\documentclass[aps,
prl,
twocolumn,
superscriptaddress,
showpacs,
amsmath,
graphicx,
amsfonts,
amssymb,
multirow,
reprint,
nofootinbib,
nobalancelastpage
]{revtex4-1}
\usepackage{graphicx}
\usepackage{siunitx}
\usepackage{latexsym}
\usepackage{amsmath}
\usepackage{txfonts}
\usepackage{helvet}
\usepackage{braket}
\usepackage{amscd}
\usepackage{longtable}
\usepackage{epsfig}
\usepackage{lipsum}

\usepackage{enumitem}
\setitemize{noitemsep,topsep=0pt,parsep=0pt,partopsep=0pt,leftmargin=*}
\setenumerate{noitemsep,topsep=4pt,parsep=0pt,partopsep=0pt,leftmargin=*}

\begin{document}

\title{Zero- to Ultralow-Field Nuclear Magnetic Resonance $J$-Spectroscopy with Commercial Atomic Magnetometers}

\author{John W. Blanchard}
\email{blanchard@uni-mainz.de}
\affiliation{Helmholtz-Institut Mainz, 55099 Mainz, Germany}

\author{Teng Wu}
%\email{teng@uni-mainz.de}
\affiliation{Helmholtz-Institut Mainz, 55099 Mainz, Germany}
\affiliation{Johannes Gutenberg-Universit{\"a}t  Mainz, 55099 Mainz, Germany}

\author{James Eills}
\affiliation{Helmholtz-Institut Mainz, 55099 Mainz, Germany}
\affiliation{Johannes Gutenberg-Universit{\"a}t  Mainz, 55099 Mainz, Germany}

\author{Yinan Hu}
\affiliation{Helmholtz-Institut Mainz, 55099 Mainz, Germany}
\affiliation{Johannes Gutenberg-Universit{\"a}t  Mainz, 55099 Mainz, Germany}

\author{Dmitry Budker}
\affiliation{Helmholtz-Institut Mainz, 55099 Mainz, Germany}
\affiliation{Johannes Gutenberg-Universit{\"a}t  Mainz, 55099 Mainz, Germany}
\affiliation{Department of Physics, University of California, Berkeley, CA 94720-7300 USA}

\date{\today}

\begin{abstract}

Zero- to ultralow-field nuclear magnetic resonance (ZULF NMR) is an alternative spectroscopic method to high-field NMR, in which samples are studied in the absence of a large magnetic field. 
Unfortunately, there is a large barrier to entry for many groups, because operating the optical magnetometers needed for signal detection requires some expertise in atomic physics and optics. 
Commercially available magnetometers offer a solution to this problem. 
Here we describe a simple ZULF NMR configuration employing commercial magnetometers, and demonstrate sufficient functionality to measure samples with nuclear spins prepolarized in a permanent magnet or initialized using parahydrogen. 
This opens the possibility for other groups to use ZULF NMR, which provides a means to study complex materials without magnetic susceptibility-induced line broadening, and to observe samples through conductive materials.
%Abstract goes here.
%[ZULF NMR is good, but there exists a substantial barrier to entry for many groups]
%[Commercially available magnetometers can make things much easier]
%[We demonstrate sufficient functionality, and describe a simple ``open-source'' configuration]
%[Broad access to ZULF NMR opens up potential applications to studies of hyperpolarization level-crossing dynamics, etc.]
%\lipsum[1]
\end{abstract}


\maketitle

\section{Introduction}

 Zero- to ultralow-field nuclear magnetic resonance (ZULF NMR) is an emerging alternative magnetic resonance modality where measurements are performed in the absence of an applied magnetic field \cite{eMagRes-ZULF}. 
%Unlike conventional NMR, in which the primary interaction is the coupling of nuclear spins to a large applied magnetic field, ZULF NMR presents a regime dominated by `local' spin-spin couplings. 
%; the Zeeman interaction is negligible or small enough that it can be treated as a perturbation on the $J$-coupling and/or dipole-dipole interactions. 
%These local interactions are sensitive to subtle changes in geometry, conformation, and electronic structure, thus serving as a valuable source of chemical information.
By eliminating the need for a large magnetic field to encode chemical information in the form of chemical shifts, ZULF NMR avoids some problems encountered by conventional NMR, such as broadening from susceptibility gradients in complex materials \cite{Tayler2018}, limited rf penetration into conductive samples \cite{SQUID-Pepper-Can, Tayler2019Metal}, and truncation of nuclear spin interactions that do not commute with the Zeeman interaction \cite{Blanchard2015,King2017}.

%In this regime, the internal spin-spin coupling Hamiltonians are not truncated by the imposed symmetry of a large magnetic field, so all information encoded in the interaction tensors is preserved [cite Blanchard2015]. 
Furthermore, the high absolute field homogeneity and the existence of decoherence-protected multiple-spin states at zero magnetic field contribute to long spin coherence times that enable high-precision measurement of nuclear spin couplings.
This has made ZULF NMR a useful tool for fundamental physics experiments searching for dark matter \cite{CASPEr-Comagnetometer} and exotic spin couplings \cite{Wu2018}.
%\item High-precision spectroscopy
%\item Can do things high-field NMR can't do (untruncated interactions, $J_{\rm anti}$)
ZULF NMR is also useful for the study and development of hyperpolarization methods \cite{Theis2011,Theis2012,Theis2015,Eills2019}, which dramatically increase the sensitivity of NMR and MRI.%, allowing for metabolic imaging [ref], surface-selective

Whereas nuclear quadrupole resonance can be detected at zero field using tuned LC circuits, signals arising from direct and indirect dipole-dipole couplings occur at much lower frequencies. 
Indirect point-by-point sampling of zero-field spin dynamics via field cycling \cite{Weitekamp1983,Zax1985} is one option with some enduring appeal \cite{Zhukov2018}, but it is often too slow for applications.
Direct detection of ZULF NMR signals requires alternative non-inductive detection modalities, which serves as a barrier to entry for many researchers.
For example, superconducting quantum interference devices (SQUIDs) have been used to detect NMR at sub-$\mu$T fields \cite{McDermott2002} and are commercially available, but the need for cryogenic temperatures and complex coil design have inhibited their widespread use.
Nitrogen-vacancy centers in diamond \cite{Devience2015,Qdyne,Kehayias2017,Smits2019} might one day prove useful for ZULF NMR, but they are not yet competitive with respect to sensitivity.

In recent years, atomic magnetomers \cite{Budker2007,romalisSERF} have emerged as the preferred detectors for ZULF NMR spectroscopy of liquid-state samples, but the design, construction, and operation of appropriate instrumentation has so far generally required substantial atomic/optical physics expertise.
Recently, however, standalone optically pumped atomic magnetometers with magnetic-field sensitivity within an order of magnitude of that achieved with state-of-the-art instrumentation have become commercially available.
One example is the QuSpin Zero-Field Magnetometer (QZFM) from QuSpin Inc. \cite{osborne2018fully}, which is based on changes in atomic absorption at a zero-field resonance \cite{DUPONTROC1969}.
These sensors have found applications in fetal magnetocardiography \cite{Batie2017}, development of wearable magnetoencephalography systems \cite{Boto2018}, trace detection of magnetic nanoparticles in complex fluids \cite{Bougas2018}, and, when paired with flux concentrators, sensitive magnetic microscopy \cite{Kim2016}.
%[List of other things QuSpins have been used for, including Savukov with his flux concentrator \cite{Kim2016}, fetal magnetocardiography \cite{Batie2017}, Lykourgos measuring magnetic particles \cite{Bougas2018}, wearable magnetoencephalography systems \cite{Boto2018}]. 
%\begin{itemize}
%\item ZULF NMR has found new applications following the use of atomic magnetometry
%\item Until now, only magnetometry experts can do ZULF with atomic magnetometers
%\item Commercial OPMs are now available (hooray QuSpin)
%\end{itemize}
Lee et al previously used an earlier version of the QuSpin QZFM magnetometer to measure low-field ($\sim$325\,nT) spin precession of deionized water doped with a stable nitroxide radical and polarized via Overhauser dynamic nuclear polarization at 58\,$\mu$T \cite{Lee2019}.
%[This is basically a high-field NMR experiment carried out in very low fields (too small even for chemical shifts) -- ZULF necessarily involves spin-spin couplings.
In contrast with ZULF NMR, such pseudo-high-field experiments require high-frequency rf irradiation, and at these fields have not been used to acquire spectra that contain chemical information.

\begin{figure}
	\includegraphics[width=\columnwidth]{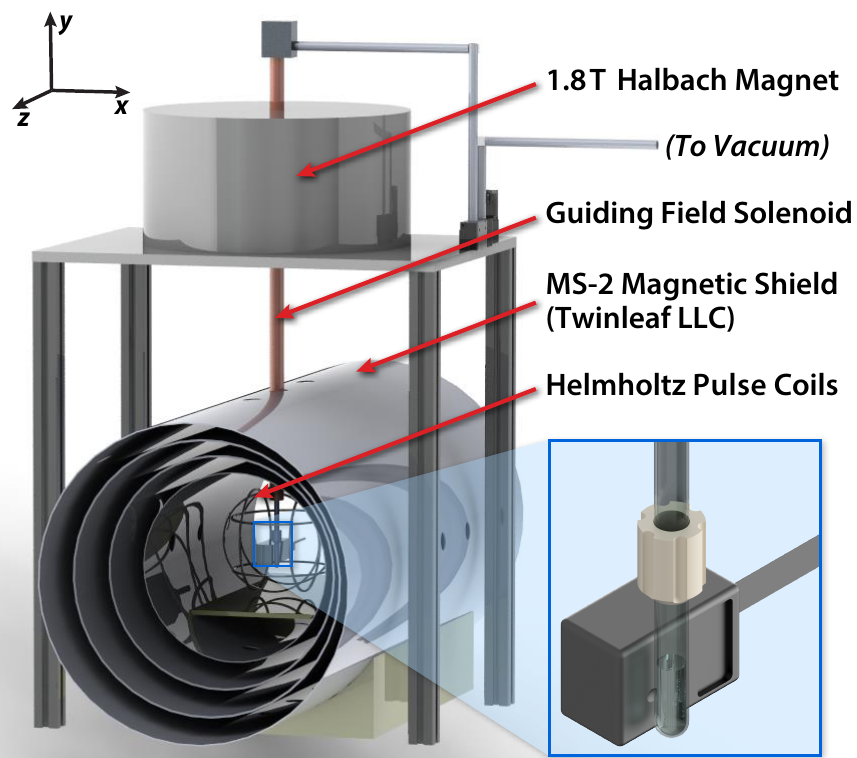}
	\caption{Experimental apparatus, as implemented for measurement of pre-polarized standard samples.
	The inset shows a magnified view of the sample and QZFM sensor.
	For experiments utilizing parahydrogen, the Halbach magnet and guiding field solenoid are removed and the sample is placed in a custom NMR tube assembly that permits bubbling of parahydrogen through the solution (see Supplemental Material).
	The end caps on each of the cylindrical magnetic shielding layers are omitted for clarity.
	%\textbf{[Note(JWB): We really need a zoomed-in part (b) showing the arrangement of the sample+sensor arrangement. Maybe we make this a full-width figure by also including the electronics layout?]}
	}
	\label{fig:schematic}
\end{figure}

Here we demonstrate the measurement of ZULF NMR $J$-spectroscopy with a second-generation QZFM magnetometer in the arrangement shown in Fig.\,\ref{fig:schematic}.
We show spectra for a set of standard samples (${}^{13}$C-formic acid, 2-${}^{13}$C-acetonitrile, and ${}^{13}$C$_2$,${}^{15}$N-acetonitrile), thermally pre-polarized in a permanent magnet before shuttling to zero field for detection. We compare performance against a state-of-the-art homebuilt ZULF spectrometer.
This arrangement is also compatible with parahydrogen-induced polarization (PHIP) in which nonequilibrium nuclear spin polarization is prepared via chemical reaction with hydrogen gas enriched in the \emph{para} spin isomer.
We show zero-field heteronuclear $J$-spectra of dimethyl maleate and %${}^{15}$N-
pyridine at natural isotopic abundance using hydrogenative and nonhydrogenative PHIP, respectively.

%\begin{itemize}
%\item We show how it works
%\item We show results for standard samples, thermally polarized
%\item We show results for samples polarized by PHIP
%\item We show results for samples polarized by SABRE
%\item We identify current limitations and suggest avenues for improvement
%\end{itemize}

\section{Results}

%\paragraph{\bf{Figures of Merit}}
%\begin{itemize}
%\item We should define these
%\item SNR for standard samples is what really counts
%\item Related to absolute sensitivity, standoff distance
%\end{itemize}

%\paragraph{Describe setup}
%\begin{itemize}
%\item Figure \ref{fig:schematic}: Experimental setup
%\end{itemize}

\begin{figure*}
	\includegraphics{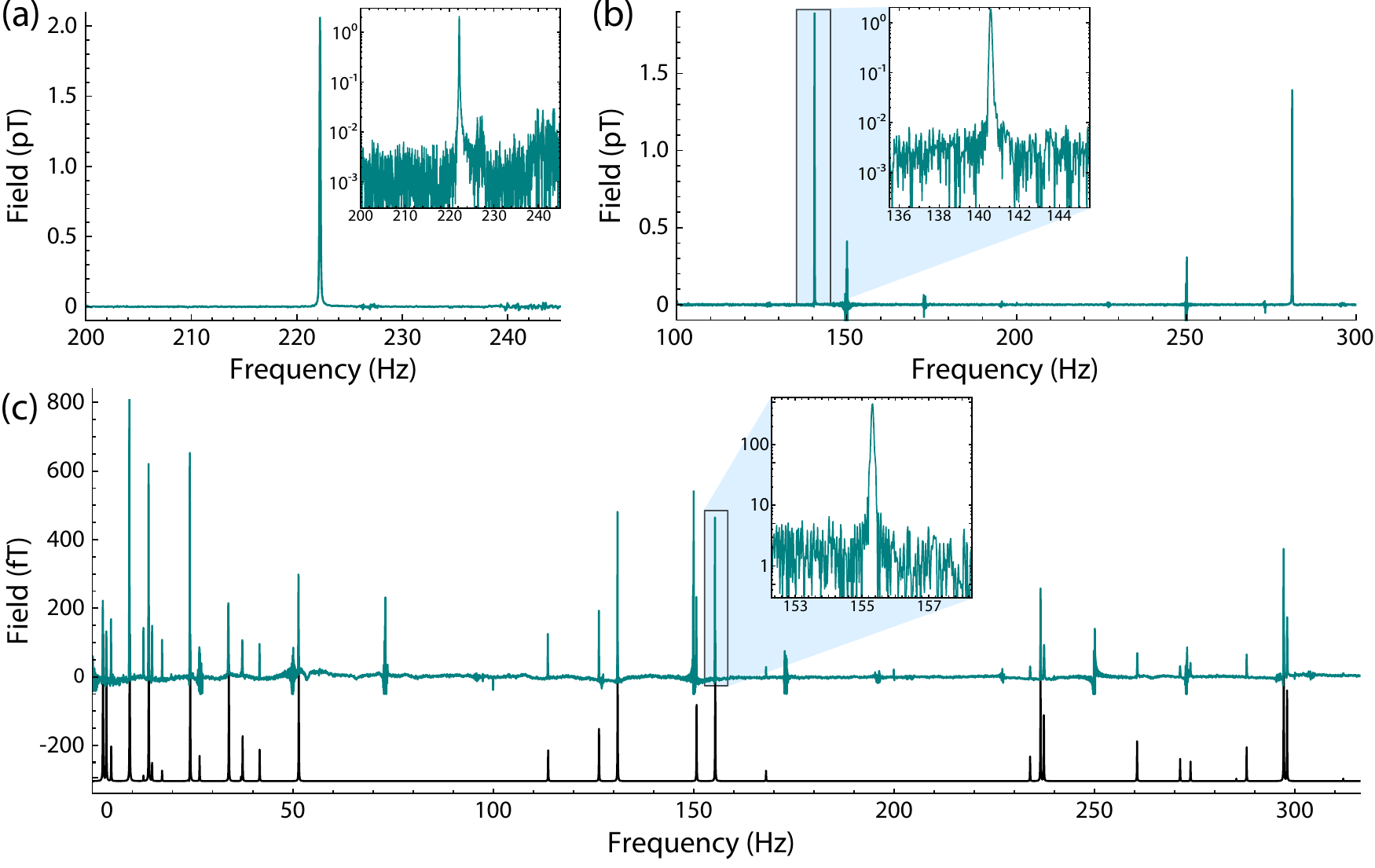}
	\caption{Spectra of ZULF NMR standards: (a) ${}^{13}$C-formic acid, (b) 2-${}^{13}$C-acetonitrile, and (c) ${}^{13}$C$_2$,${}^{15}$N-acetonitrile. 
	Part (c) includes a simulated spectrum in order to clarify which peaks are due to NMR signals.
	Each spectrum is the result of 32 averages.
	Insets show selected resonances on a logarithmic scale with the same units.
	The noise floor is consistent with the specified magnetometer sensitivity, $\lesssim15~\rm{fT}/\sqrt{\rm{Hz}}$.
	Noise at $100 n \pm 27~\rm{Hz}~(n=0,1,2,...)$ is suspected to arise due to higher overtones of the 50\,Hz line noise interfering with a 923\,Hz internal modulation of the QZFM sensors.
	}
	\label{fig:Standards}
\end{figure*}

\paragraph{\bf{Spectra of pre-polarized standard samples}}
In order to evaluate the performance of the QZFM sensors as detectors for ZULF NMR, we performed measurements on a set of standard samples via the following procedure:
\begin{enumerate}
\item Following polarization at 1.8\,T for 20\,s, 
the sample is shuttled into the magnetically shielded detection region.
\item While shuttling, a guiding magnetic field is applied using a solenoid wrapped around the shuttling tube, as well as the $x$-axis 
Helmholtz pulsing coil.
\item After the sample arrives next to the sensor, the solenoid current is turned off adiabatically.
\item The $x$-axis 
pulse coil current is then turned off suddenly ($< 10 ~\mu\rm{s}$), and a pulse\footnote{As the sudden transfer to zero field is sufficient to induce an oscillating signal, the final pulse is technically optional, but frequently provides an enhancement by swapping $I_z+S_z$ and $I_z-S_z$ spin order.} is applied along the $z$ axis with area $\gamma_C B_p \tau = \pi$, where $\gamma_C$ is the $^{13}$C gyromagnetic ratio, $B_p$ is the pulse amplitude, and $\tau$ is the pulse duration ($50~\mu\rm{s}$ for all experiments in this work).
\item Immediately following the pulse, the magnetic signal produced by the sample along the $x$ axis 
is measured via the QZFM magnetometer (the analog output of the sensor is read out by a National Instruments NI~9239 analog input card).
\end{enumerate}

The spectrum of ${}^{13}$C-formic acid, a heteronuclear two-spin system (the acidic proton can be neglected due to fast exchange) is shown in Fig.\,\ref{fig:Standards}(a).
As has been explained in Ref. \cite{Ledbetter2011}, at zero magnetic field there is one observable nuclear spin transition, which occurs at the $J$-coupling frequency ${}^1J_{\rm{CH}} \approx 222$\,Hz. 
The signal-to-noise ratio (SNR) after 32 averages is $\sim 500$. %[The single-scan SNR is $\sim XXX$.]
For a state-of-the-art instrument such as was used in Refs.\,\cite{CASPEr-Comagnetometer,Billion2018}, the single-scan SNR for the same sample may be as high as 750.

Figure\,\ref{fig:Standards}(b) shows the zero-field nuclear spins resonances of 2-${}^{13}$C-acetonitrile, a heteronuclear four-spin system composed of one ${}^{13}$C and three equivalent ${}^{1}$H nuclei, with a one-bond coupling ${}^1J_{\rm{CH}} \approx 141$\,Hz (the ${}^{14}$N nucleus can be neglected due to self-decoupling via fast quadrupolar relaxation). 
As explained in Refs. \cite{Butler1,Theis2013}, the zero-field spectrum consists of two peaks at ${}^1J_{\rm{CH}}$ and $2\times {}^1J_{\rm{CH}}$.
The SNR of the peak at ${}^1J_{\rm{CH}}$ after 32 averages is $\sim 350$. %[The single-scan SNR is $\sim XX$.]

The spectrum of ${}^{13}$C$_2$,${}^{15}$N-acetonitrile is shown in Fig.\,\ref{fig:Standards}(c). 
This strongly coupled six-spin system yields a larger number of peaks, spread over the 0--300\,Hz spectral range.
A simulated spectrum is shown below the experimental spectrum in order to clarify which peaks correspond to NMR signals and which correspond to environmental/electronic noise.
The SNR of the peak at 155.3\,Hz after 32 averages is $\sim 80$.
%A number of peaks originating due to noise are visible, and can be identified based on the spectrum (shown below in red) collected in the absence of a sample.

For all spectra in Fig.\,\ref{fig:Standards}, the linewidths are limited by the Fourier resolution: 0.1\,Hz for 10\,s acquitions.
There is no evidence that nuclear spin coherence times are affected by the magnetometer. 

%\paragraph{Standard 1: ${}^{13}$C-formic acid}
%\begin{itemize}
%\item Figure \ref{fig:Standards}A: Formic acid spectrum
%\item Everyone's favorite sample
%\item SNR around 220 Hz
%\end{itemize}

%\paragraph{Standard 2: 2-${}^{13}$C-acetonitrile}
%\begin{itemize}
%\item Figure \ref{fig:Standards}B: Acetonitrile spectrum
%\item SNR around 140 Hz and 280 Hz
%\end{itemize}

%\paragraph{Standard 3: ${}^{13}$C$_2$,${}^{15}$N-acetonitrile}
%\begin{itemize}
%\item Figure \ref{fig:Standards}C: Fully-labeled acetonitrile
%\item Peaks everywhere!
%\end{itemize}

\begin{figure}
	\includegraphics[width=\columnwidth]{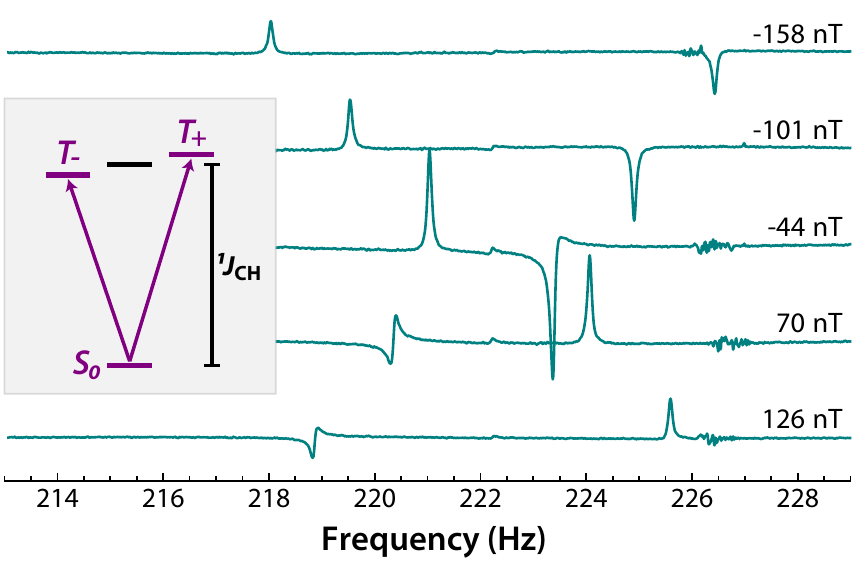}
	\caption{Spectra of ${}^{13}$C-formic acid vs. applied magnetic field. 
	Inset shows nuclear spin energy levels and transitions. 
	The phase of each spectrum is adjusted such that the $|S_0\rangle \leftrightarrow |T_+\rangle$ transition is proportional to an absorptive Lorentzian.
	}
	\label{fig:NZF}
\end{figure}

\paragraph{\bf{Operation in non-zero magnetic fields}}
Application of a small magnetic field increases the amount of information that can be extracted from ZULF NMR spectra \cite{Ledbetter2011}.
However, sensitive magnetometers such as those operating in the spin-exchange relaxation-free (SERF) regime are optimized for operation at zero magnetic field, and performance typically degrades in the presence of larger magnetic fields \cite{Ledbetter2008Cs}.

To evaluate the sensitivity of the QZFM sensor as a function of applied magnetic field, a series of ${}^{13}$C-formic acid spectra were acquired in the presence of an ultra-low magnetic field of varied intensity. 
The results are shown in Fig.\,\ref{fig:NZF}. 
The field was applied in the $y$ direction, orthogonal to the sensitive axes of the magnetometer. 
This induces an additional peak splitting, as the magnetic field lifts the degeneracy between the heteronuclear spin-triplet energy levels.

%[SNR decreases at higher fields]
The maximum measured signal amplitude in Fig.\,\ref{fig:NZF} is $\sim$\,840~fT, obtained with an applied field of -44\,nT.
The decrease in sensitivity as a function of applied field is consistent with a Lorentzian profile having a $\sim$100\,nT linewidth (full width at half height), presumably related to the width of the atomic zero-field resonance.
%[SNR drops to (cutoff) at XX nT]
This is also consistent with QuSpin Inc.'s suggestion to operate their sensors in ambient magnetic fields smaller than 50\,nT.
%[Maybe SNR improves again at higher fields on the other side of the Rb resonance?]

%\paragraph{Gradiometer configuration}
%\begin{itemize}
%\item Figure 3
%\item general SNR improvement?
%\item common-mode noise rejection
%\end{itemize}

%\paragraph{Noise analysis}
%\begin{itemize}
%\item Figure 4?
%\end{itemize}

\begin{figure}
	\includegraphics[width=\columnwidth]{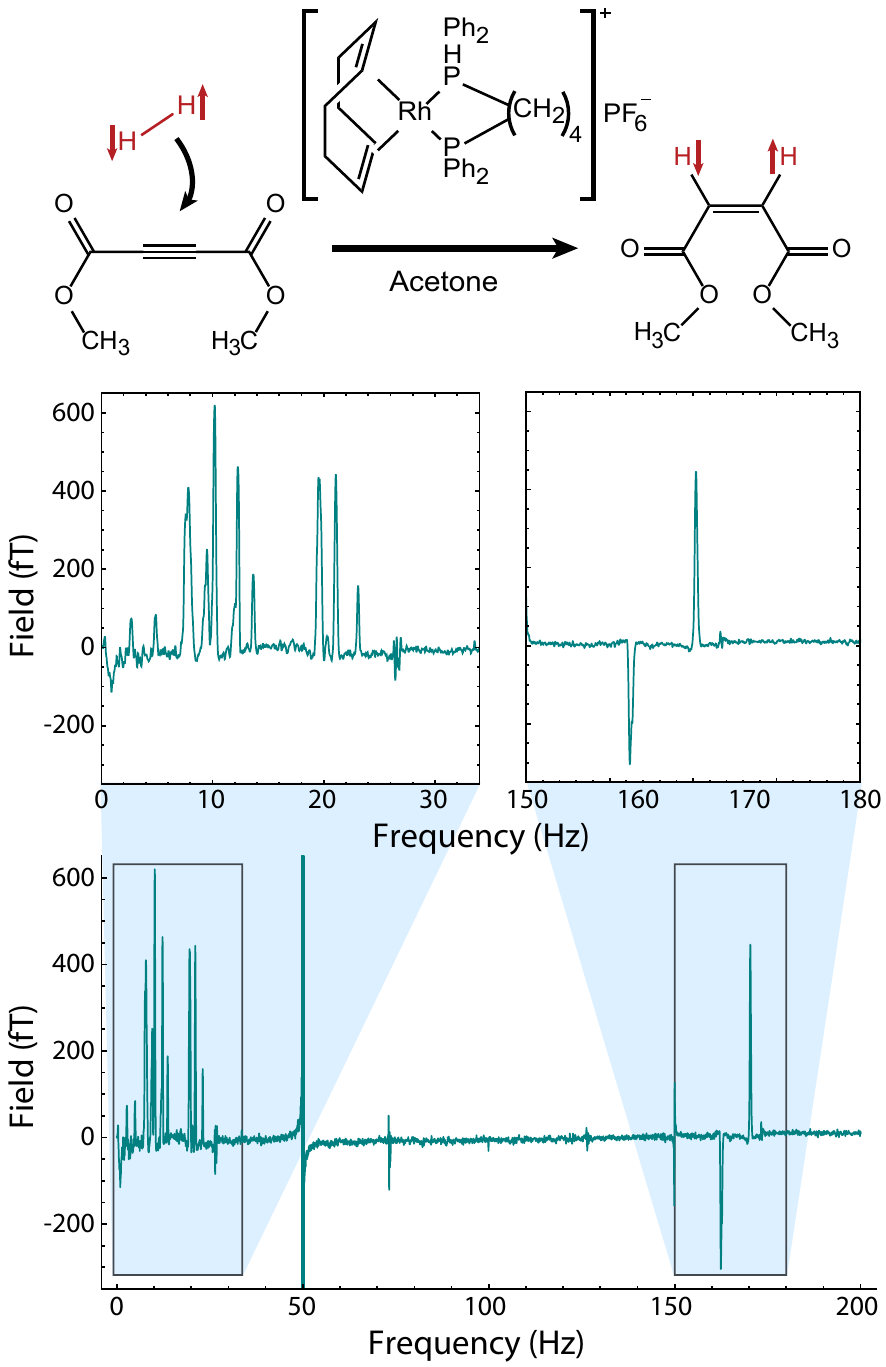}
	\caption{Spectrum of dimethyl maleate at natural ${}^{13}$C abundance, polarized by addition of parahydrogen to dimethylacetylene dicarboxylate via the reaction shown at the top of the figure.}
	\label{fig:PHIP}
\end{figure}

\paragraph{\bf{Samples polarized via parahydrogen}}
Parahydrogen-induced polarization can produce dramatically enhanced NMR signals, and ZULF NMR can be used to study the relevant/necessary spin dynamics. In Fig.\,\ref{fig:PHIP}, a ZULF NMR spectrum of a PHIP reaction mixture is shown (chemical reaction is shown in the inset). 
The 2\% natural abundance of 2-$^{13}$C-dimethyl maleate produces characteristic antiphase peaks centered around 167 Hz, which corresponds to the ${}^1J_{\rm{CH}}$ coupling constant \cite{Theis2011}, as well as peaks at low frequency.
The 2\% natural abundance of 1-$^{13}$C-dimethyl maleate produces a spectrum with all peaks below 25 Hz, because this isotopomer contains no 1-bond $J_{\rm{CH}}$ couplings.
 
%ZULF NMR can be utilized to study how the resulting nuclear-spin magnetization depends on magnetic field, and to exploit avoided crossings of nuclear-spin energy levels that necessarily occur near zero field to do nice things [cite James JCP 2019].
%\begin{itemize}
%\item Figure \ref{fig:PHIP}
%\item Sample: dimethylacetylene dicarboxylate
%\item Woo, signal
%\end{itemize}

\begin{figure}
	\includegraphics[width=\columnwidth]{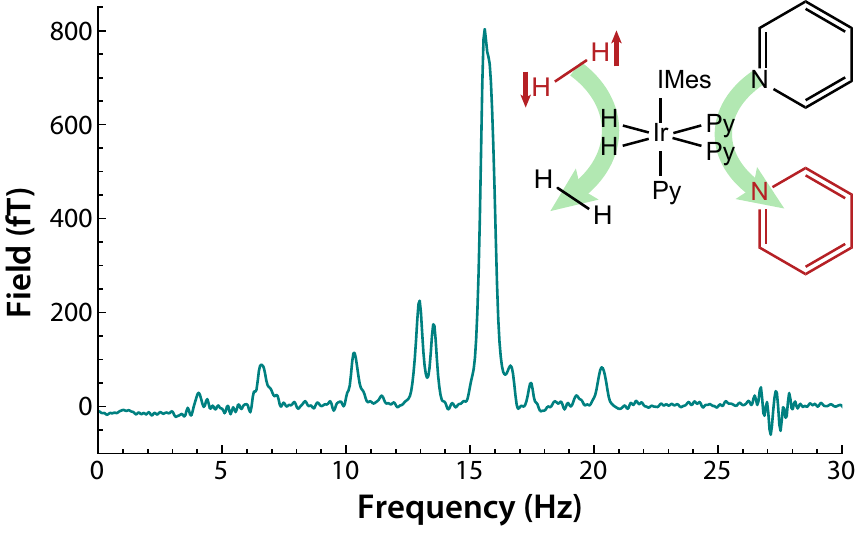}
	\caption{Spectrum of ${}^{15}$N-pyridine at natural isotopic abundance polarized via signal amplification by reversible exchange (SABRE). 
	The SABRE reaction scheme, involving reversible exchange of both parahydrogen and pyridine with the iridium polarization transfer complex, is shown inset.
	The signal at 27 Hz arises due to electronic noise from the QZFM.}
	\label{fig:SABRE}
\end{figure}

It is also possible to use parahydrogen to hyperpolarize molecules without the need for chemical addition of hydrogen, via SABRE \cite{Adams2009}.
Figure\,\ref{fig:SABRE} shows a ZULF NMR spectrum of ${}^{15}$N-pyridine (${}^{15}$N present in natural isotopic abundance) spin polarized via SABRE at zero field.
The reversible exchange of parahydrogen and pyridine with the iridium complex is shown in the inset.

%\begin{itemize}
%\item Figure \ref{fig:SABRE}
%\item Sample: ${}^{15}$N-pyridine
%\item Woo, signal
%\end{itemize}

\section{Discussion}

Commercial optically pumped magnetometers such as the QZFM offer a route to the detection of ZULF NMR for researchers who are not experts in atomic physics, as they are class-1 laser devices operating near physiological temperatures with a small footprint ($\sim$5\,cm$^3$). % and relatively low cost ($\leq 10\rm{k}$~USD). 
The SNR achieved using the QZFM is within an order of magnitude of state-of-the-art instruments.
This room-temperature operation is particularly useful for the detection of samples polarized via chemical reaction with parahydrogen, because the chemical kinetics are highly temperature dependent. Unlike previous zero-field measurements of SABRE, no sample cooling is required. This is particularly useful for studying volatile or unstable compounds, or compounds dissolved in low-boiling-point solvents.

The QZFM sensors do, however, have a number of drawbacks.
%[overheating]
These second-generation QZFM sensors suffer from laser overheating when used in confined geometries, especially when multiple sensors are employed.
We have found that this issue can be mitigated by flowing air over the sensor, but this could become problematic for experiments at elevated temperature.
%[limited bandwidth]
The sensors also have a limited bandwidth due to a low-pass %hardware 
filter at 500\,Hz, which precludes measurement of molecules with larger $J$-couplings, such as those with a direct P-H or P-F bond.

Many substantial noise peaks appear in spectra collected with these sensors.
Those occurring at multiples of 50\,Hz (European power line frequency) are due to electronic noise in the coils.
%[Ask Vishal about the other ones at 27, 73, 127, 173, 227, 273 Hz]
Noise at frequencies of %$100 n \pm 27~\rm{Hz}$
27, 73, 127, 173, ... Hz may be due to interference between higher overtones of the line frequency and the 923\,Hz modulation frequency used in the QZFM sensors.
Common-mode noise can be removed by using two or more sensors in a gradiometric configuration \cite{ZULF-Grad}.

Sensitivity is attenuated at magnetic fields higher than $\sim$\,50\,nT.
While it is possible to re-optimize a homebuilt magnetometer to operate at different magnetic fields, %these commercial sensors offer little control or flexibility in their operation.
the only remedy available for these commercial sensors is the use of built-in field-cancellation coils, but these have the downside of also applying an inhomogeneous magnetic field to the NMR sample.

The response of the magnetometers to sudden field changes remains problematic, especially when the field is on for longer than 1\,ms.
When the field is on for longer times before being switched off, the feedback system used to stabilize the temperature of the vapor cell is affected, resulting in long-term low-frequency fluctuations in the sensor output. These fluctuations can prove problematic for data processing, especially for samples exhibiting low-frequency resonances.
For short pulses, such as the 50\,$\mu$s pulses used here, the QZFM response is not significantly longer than that of a state-of-the-art homebuilt instrument.

While we have used a sensor sold by QuSpin Inc. for these measurements, there are a number of commercial vendors (e.g. Twinleaf LLC) that supply atomic magnetometers potentially suitable for ZULF NMR applications.

\paragraph{\bf{Outlook}}
ZULF NMR allows for the study of systems inaccessible by regular high-field NMR.
Applications include in situ optimization of nuclear spin hyperpolarization methods such as SABRE-SHEATH \cite{Theis2011,Theis2012,Theis2015} and field-dependent spin polarization transfer experiments \cite{Eills2019}, as well as general studies of nuclear spin dynamics in ultralow magnetic fields \cite{Tayler2019}.
Further applications include measurement of samples confined in porous \cite{Tayler2018} or magnetic materials \cite{Tayler20192}.
The availability of commercial magnetometers appropriate for ZULF NMR detection dramatically reduces the main barrier to entry into the growing field of ZULF NMR spectroscopy. 

%[Something about anticipated applications, e.g. SABRE-SHEATH, polarization transfer, ULF relaxometry, study of heterogeneous samples]
Standalone commercial magnetometers also afford additional flexibility in the construction of ZULF NMR experiments.
Previous homebuilt ZULF NMR detectors have generally needed to measure samples from below, but sensors like the QZFM can just as easily be placed to the side of the sample.
The sensors can, for example, be placed outside of a piercing solenoid, allowing for magnetic fields to applied to the nuclear spins without affecting the sensors \cite{Xu2006}.
Such an arrangement may prove advantageous for experiments searching for a nuclear gravitational dipole moment \cite{Kimball2017, Wu2018}, and may possibly enable operation of a self-oscillating nuclear spin magnetometer \cite{Suefke2017,BillionMaser}.
Larger sensor arrays are also readily constructed, providing common-mode noise rejection \cite{ZULF-Grad}, and spatial resolution \cite{Boto2018}.

\section{Methods}

\footnotesize

\paragraph{\footnotesize{\bf{Experimental}}} %\textbf{[Note(JWB): Do we actually need this?]} 
%Apparatus/Construction + control/acquisition
The ZULF NMR apparatus (as shown in Fig.\,\ref{fig:schematic} and in further detail in the Supplemental Material) is based on a commercial optically pumped magnetometer (QZFM, QuSpin, Inc.) placed in a 3D-printed holder.
The printed holder also serves as a former for the three orthogonal Helmholtz ``pulse'' coils.
The magnetometer and pulse coil assembly is centered within a four-layer mu-metal magnetic shield (MS-2, Twinleaf LLC).
The magnetometer is oriented such that the two sensitive axes are along the $x$ and $z$ axes shown in Fig.\,\ref{fig:schematic}.
The distance between the center of the sample and the center of the magnetometer cell is 9.5\,mm.

The analog outputs of the magnetometer were read out by a National Instruments NI~9239 analog input card at 5000 samples/s.
Typically, only the projection of the magnetic field along the $x$ axis was recorded, but there are certainly applications where measurement of the correlated signals along $x$ and $z$ is advantageous (e.g., for precessing magnetization).

Background magnetic fields were controlled via a set of coils built in to the magnetic shield; the currents in the $B_x$, $B_y$, and $B_z$ coils were provided by Krohn-Hite Model 523 DC current sources (alternative stable current sources, such as those provided by Twinleaf LLC or Magnicon GmbH, are also suitable).
Fields were set to zero by minimizing the Zeeman splitting in the spectrum of ${}^{13}$C-formic acid (see Fig.\,\ref{fig:NZF}).

A Kea2 NMR console (Magritek Ltd.) with Gradient Driver Module was used for control of experimental timing (using TTL outputs) and magnetic field pulse generation (using the analog output of the gradient module).
In principle, experimental control can be achieved using any system with digital timing and analog output capabilities (see, for example, Refs.\,\cite{ZULF-RSI,Tayler2018}).

\paragraph{\footnotesize{\bf{Standard Samples}}}
%[Acquired from Sigma]
${}^{13}$C-formic acid, 2-${}^{13}$C-acetonitrile, and ${}^{13}$C$_2$,${}^{15}$N-acetonitrile were obtained from Isotec Stable Isotopes (Sigma-Aldrich)
and degassed via several freeze-pump-thaw cycles to remove any dissolved oxygen 
before being flame-sealed under vacuum ($<10^{-6}$\,bar above the frozen sample).

\paragraph{\footnotesize{\bf{Parahydrogen}}} To generate para-enriched hydrogen gas% at 95\% para enrichment
, regular hydrogen gas (purity $>99.99\%$) was passed over a hydrated iron(III) oxide catalyst (30-50 mesh, Sigma-Aldrich, Taufkirchen) at 25\,K.

For hydrogenative PHIP experiments, the initial solution was 5\,mM 1,4-bis(diphenylphosphino)butane(1,5-cyclooctadiene)rhodium tetrafluoroborate and 150\,mM dimethyl acetylenedicarboxylate in 500\,$\mu$L acetone. 
For nonhydrogenative PHIP experiments, the initial solution was 25\,mM 1,3-bis(2,4,6-trimethylphenyl)imidazole-2-ylidene(1,5-cyclooctadiene)iridium chloride and 2\,M pyridine in 300\,$\mu$L methanol.
Parahydrogen was bubbled through this solution at a pressure of 5\,bar until it became transparent (a few minutes), indicating the catalyst was fully activated.

To acquire hyperpolarized spectra, parahydrogen was bubbled into each solution for 8\,s at 5\,bar, followed by a magnetic field pulse and signal acquisition. The pulse was a 117 \,$\mu$T field applied along the detection axis for 50\,$\mu$s, which corresponds to a $\pi/2$ rotation of the proton spins.

\paragraph{\footnotesize{\bf{Signal Processing}}} All signal processing was performed using Wolfram Mathematica \cite{wolfram1991mathematica}.
To account for the magnetometer response to the magnetic field pulse, the first 50-60\,ms of data was dropped and reconstructed via backward prediction.
Long-term background drifts in the signal were then removed by subtracting a moving average from the data.
Optional exponential apodization was performed by multiplying the signal with a decaying exponential, and digital resolution was increased by zero filling to a total length four times that of the original data.
Linear phase correction was applied to the complex Fourier-transformed data using parameters that provided a fully in-phase spectrum of ${}^{13}$C$_2$,${}^{15}$N-acetonitrile.

\bibliography{QuSpin-ZULF}

\begin{acknowledgments}
The authors would like to thank Dr. Arne Wickenbrock (Uni-Mainz) and Dr. Vishal Shaw (QuSpin Inc.) for support and useful discussions.
J.E. has received funding from the European Union's Horizon 2020 research and innovation programme under the Marie Sk{\l}odowska-Curie Grant Agreement No. 766402.
Y.H acknowledges funding from the Deutsche Forschungsgemeinschaft (DFG) through the DIP program FO 703/2-1.
DB was supported in part by the U.S. National Science Foundation (CHE-1709944), 
by the European Research Council (ERC) under the European Union Horizon 2020 research and innovation program (grant agreement No 695405),
by the German Federal Ministry of Education and Research (BMBF) within the Quantumtechnologien program (FKZ 13N14439 and FKZ 13N15064),
by the DFG via the Reinhart Koselleck project, 
and by the Cluster of Excellence Precision Physics, Fundamental Interactions, and Structure of Matter (PRISMA+ EXC 2118/1) funded by the DFG within the German Excellence Strategy (Project ID 39083149).

\end{acknowledgments}

\end{document}